\documentclass[a4paper,11pt]{article}
\usepackage[utf8]{inputenc}
\usepackage{authblk}

\pdfoutput=1

\usepackage{times}
\usepackage{bm}
\usepackage{natbib}
\setcitestyle{authoryear}
\usepackage{tikz}
\usepackage{amsmath,amssymb}

\newtheorem{definition}{Definition}
\newtheorem{example}{Example}
\newtheorem{lemma}{Lemma}

\newtheorem{theorem}{Theorem}
\newtheorem{proof}{Proof}

\usepackage{geometry}
\usepackage{color,soul}

\begin{document}
\date{}
\title{Local and global robustness in conjugate Bayesian analysis} 
\author{Vahed Maroufy  \hspace{1cm}{\it and}\hspace{1cm}  Paul Marriott}
\affil{\small Department of Statistics and Actuarial Science, University of Waterloo}
\maketitle

\begin{abstract}
This paper studies the influence of  perturbations of conjugate priors in  Bayesian inference. A  perturbed prior is defined
inside a larger family,  local mixture models, and the effect on posterior inference is studied. 
The perturbation, in some sense, generalizes the linear perturbation studied in \cite{Gustafson1996}. It
is intuitive, naturally normalized and is  flexible for statistical applications. Both global and local
sensitivity analyses are considered. A geometric approach is employed for optimizing the sensitivity direction
function, the difference between posterior means and the  divergence function between posterior predictive models. 
All the sensitivity measure functions are defined on a
convex space with non-trivial boundary which is shown to be a smooth manifold.
\end{abstract}

{\it Keywords}: Bayesian sensitivity; Local mixture model; Perturbation space; Newton's method; Smooth manifold.

\section{Introduction}\label{Introduction}
Statistical analyses are often performed under certain assumptions which  are not directly
validated. Hence, there is always  interest in  investigating the degree
to which a statistical inference is sensitive to perturbations of the model and data. Specifically, in a Bayesian analysis for which conjugate priors have been chosen the  sensitivity of the posterior to prior choice is an important issue. A rich literature on sensitivity  to  perturbations of data, prior  and sampling distribution
exists in
\citet*{Cook1986}, \citet*{Mcculloch1989}, \cite{Lavine1991}, \citet*{Ruggeri1993}, \citet*{Blyth1994}, \citet*{Gustafson1996},
\citet*{Critchley2004}, \citet*{Linde2007} and \citet*{Zhu2011}.

Sensitivity analysis with respect to a  perturbation of the prior, which is the focus of this paper,
is commonly called robustness
analysis. A comprehensive literature and review of  existing methods can be found in \cite{Insua2000}.
In robustness analysis it is customary to choose a base prior model and a plausible class of perturbations.
The influence of a perturbation is assessed either locally, or globally, by measuring the divergence of
certain features of the posterior distribution. For instance,
\citet*{Gustafson1996} studies linear and non-linear model perturbations, and \citet*{Weiss1996} uses a multiplicative
perturbation to the base prior and specifies the important perturbations using the posterior
density of the parameter of interest.  Common global measures of influence include  divergence functions \citep{Weiss1996}
and relative sensitivity \citep{Ruggeri2000}. Note that  any   analysis  highly depend on the selected influence measure, see in particular 
\cite{Sivaganesan2000}.  

In local analysis,  the rate at which a posterior quantity changes, relative to the prior, quantifies sensitivity
(\citealp{Gustafson1996}; \citealp{Linde2007}; \citealp{Berger2000}). \cite{Gustafson1996}, which we follow closely,
obtains the direction in which a certain posterior expectation has the maximum sensitivity to prior perturbation by
considering a mapping from the space of perturbations to the space of posterior expectations.  In \citet*{Linde2007},
the Kullback-Leibler and $\chi^2$ divergence functions are utilized for assessing local sensitivity with respect to a
multiplicative perturbation of the base prior or likelihood model. They approximate the local sensitivity using the
Fisher information of the mixing parameter in additive and geometric mixing.

In this paper we consider both local, and global, sensitivity analyses with respect to   perturbations of a conjugate
base prior.  We aim for three important properties for our method. Firstly, a well-defined perturbation space whose
structure is such that it   allows  the analyst to select the generality of the perturbation in a clear way. Secondly,
we want the space to be tractable, hence we look at convex sets inside linear spaces. Finally we want, in order to
allow for meaningful comparisons, the space to be  consistent with elicited  prior knowledge. Thus if a subject
matter expert indicates  a prior moment or quantile  has a known value -- or if a constraint such as symmetry  is
appropriate --  then all perturbed priors should be consistent with respect to this information.

Such an approach to defining the  perturbation space extends  the linear perturbations studied in \cite{Gustafson1996}
in all three ways. We do not require the same positivity  condition, rather use one which is more  general and returns
naturally normalized distributions.  Further, our  space is highly tractable, due to intrinsic linearity and convexity.
Finally it is clear, with our formulation, how to remain consistent with prior information which may have been elicited
from an expert.  The cost associated with this generalisation  is the boundary defined in by (\ref{Poss condition}) in
\S \ref{Theory and Geometry} and the methods we have developed to work with it.   We also can compare our method with the
geometric approach of \cite{Zhu2011}  which uses a manifold based approach. Our, more linear,  approach considerably improves
interpretability and tractability while sharing  an underlying geometric foundation. 

In the examples of this paper we work with our perturbation space in three ways. Similarly to \cite{Gustafson1996} and
\cite{Zhu2011} in Example 1 we look for the worst possible perturbation, both locally and globally.  In Example 2 we add
constraints to the perturbation space, representing prior knowledge, and again look for maximally bad local and global
perturbations. Finally, in Example 3, we marginalise over the perturbation space -- rather than optimising over it -- 
as a way of dealing with the uncertainty of the prior. 

The paper is organized as follows. In Section \ref{Perturbation Space}, the perturbation space is introduced
and its properties are studied. Sections \ref{Local Sensitivity} and \ref{Global sensitivity}
develop the theory of local and global sensitivity analysis.
Section \ref{estimating lambda} describes the geometry of the perturbation parameter space
and proposes possible algorithms for quantifying local and global sensitivity. 
In Section \ref{Examples} we examine three examples. The proofs are sketched in Appendix.

\section{Perturbation Space}\label{Perturbation Space}

\subsection{Theory and Geometry}\label{Theory and Geometry}
We construct a perturbation space using the following definitions (\citealp{Marriott2002}; \citealp{Marriott2006};
\citealp{Anaya-Izquierdo2007}). For more details about convex and differential geometry see \cite{Berger1987} and
\cite{Amari1990}.

\begin{definition}\label{Pert_space}
For the family of mean parameterized models $f(x;\theta)$ the perturbation space is defined by the family of models
$f(x;\theta,\lambda)$ such that, \\
(i) $f(x;\theta,0)=f(x;\theta)$ for all $\theta$.\\
(ii) $f(x;\theta_0,\lambda)-f(x;\theta_0)$ is Fisher orthogonal to the score of $f(x;\theta)$ at $\theta_0$.\\
(iii) For fixed $\theta$ the $f(x;\theta_0,\lambda)$ space is affine in the mixture ($-1$) affine geometry defined in \citealp{Marriott2002}.
\end{definition}
A natural way to implement Definition \ref{Pert_space}  is to extend  the family  $f(x;\theta)$ by attaching to it, at each $\theta_0$, the subfamily
$f(x;\theta_0,\lambda)$, which is finite dimensional and spanned by a set of linearly independent functions $v_j(x;\theta_0)$,
$j=1,\cdots,k$, all Fisher orthogonal to the score of $f(x;\theta)$ at $\theta_0$. Thus, the subfamily $f(x;\theta_0,\lambda)$
can be defined as the linear space $f(x;\theta_0)+\sum\lambda_jv_j(x;\theta_0)$, where $\lambda_j$ is a component of the vector
$\lambda$. For $f(x;\theta_0,\lambda)$ to be a naturally normalized
density, we need two further restrictions: (i) $\int v_j dx =0$, and (ii) the $\lambda$ parameters
must be restricted such that each subfamily is non-negative for all $x$.  This defines the parameter space as 
\begin{equation}\label{Poss condition}
\Lambda_{\theta_0}=\left\{\lambda\,\,|\,\,\, f(x;\theta_0)+\sum\lambda_jv_j(x;\theta_0)\geq 0,\,\,\, \text{for all}\,\, x\right\}.
\end{equation}
Note the space $\Lambda_{\theta_0}\subset R^k$, is an  intersection of half-spaces and consequently is convex (\citealp[Ch.11]{Berger1987}). 

Clearly, to construct such a  perturbation space, the functions $\nu_j$ must be selected. A particular  form of Definition
\ref{Pert_space} with naturally specified $\nu_j$'s is the family of local mixture models. This family is introduced in
\citet*{Marriott2002} as an asymptotic approximation to a subspace of continuous mixture models with small mixing variation
relatively to the total variation.  Because of this small, or local, assumption, all perturbations are, in some sense, close to the baseline prior, and so any correspondingly large changes in the posterior will be of interest, as we show in the examples. 
\begin{definition}\label{loc_mix}
The local mixture of a regular exponential family $f(x;\theta)$ of order $k$ via its mean parameterization,
$\theta$, is defined as
\begin{eqnarray}
h(x;\lambda,\theta)=f(x;\theta)+\lambda_2 \, f^{(2)}(x;\theta) +\cdots+\lambda_k \, f^{(k)}(x;\theta), \hspace{.5cm}\lambda \in \Lambda_{\theta}\subset R^{k-1} \label{lmm_4}
\end{eqnarray}
where $\lambda=(\lambda_2,\cdots,\lambda_k) \in \Lambda_{\theta}$ and $f^{(j)}(x;\theta) =\frac{\partial^j}{\partial \theta^j}f(x;\theta)$, $(j=1,\cdots,k)$. 
Also, $\Lambda_\theta$, for any fixed and known $\theta$, is a convex space defined by a set of supporting hyperplanes. 

\end{definition}
For regular exponential family $\int f^{(j)}(x;\theta_0)\,dx=0$, and as shown in \cite{Morris1982}, for natural exponential
family, $f^{(j)}(x;\theta_0)$'s are linearly independent and all Fisher orthogonal to the score function at $\theta_0$.
This family is identifiable in all parameters, behaves locally similar to genuine mixture models, yet it is richer
in the sense that compared to a regular density function with the same mean they can also produce smaller variance. 
Further properties of these models are studied in \citet*{Anaya-Izquierdo2007}. 

\subsection{Prior Perturbation}\label{Local Perturbation}
Suppose the base prior model is $\pi_{0}(\mu;\theta)$, the probability (density) function of a natural
exponential family with the hyper-parameter $\theta$. 
\begin{definition}\label{perturb_prior}
The perturbed prior model corresponding to $\pi_{0}(\mu;\theta)$ is defined by 
\begin{eqnarray}
\pi(\mu,\lambda;\theta) &:=& \pi_{0}(\mu;\theta) + \sum\nolimits_{j=2}^{k}{\lambda_j \, \pi_{0}^{(j)}(\mu;\theta)}\nonumber\\
&=& \pi_{0}(\mu;\theta) \left\{1 + \sum\nolimits_{j=2}^{k}{\lambda_j \, q_j(\mu,\theta)}\right\}, \hspace{1cm} \lambda \in \Lambda_{\theta} \label{local_perturb2}
\end{eqnarray}
where $\lambda=(\lambda_2,\lambda_3,\cdots,\lambda_k)$ is the perturbation parameter vector, and $q_j(\mu,\theta)=\frac{\pi_{0}^{(j)}(\mu;\theta)}{\pi_{0}(\mu;\theta)}$
are polynomials of degree $j$.
\end{definition}
In Definition (\ref{perturb_prior}), $\pi_0$ is perturbed linearly, similar to the linear perturbation 
\begin{eqnarray}
\tau(\cdot,\pi_0,u^*)=\pi_0(\cdot)+u^*(\cdot), \hspace{1cm} u^*(\cdot)>0  \label{Linear perturb}
\end{eqnarray}
studied in \citet*{Gustafson1996}, but with a different positivity condition, and is, as we shall show, very interpretable for  applications.
Definition (\ref{perturb_prior}) 
can also be seen as the multiplicative perturbation model $\pi(\mu,\lambda;\theta) =\pi_0(\mu,;\theta)\,h^{*}(\mu;\lambda,\theta)$
studied in \citet*{Linde2007}.

As shown 
in \citet*{Anaya-Izquierdo2007}, the base and perturbed models share the same mean $\theta$; however, the perturbation
is implemented through changing the higher order moments by adding linear combinations of $\lambda$. This fact
grantees the properties mentions in Section \ref{Introduction}.

\section{Local Sensitivity}\label{Local Sensitivity} 
In this section we study the influence of local perturbations, defined inside the perturbation space, 
on the posterior mean. Similar to \cite{Gustafson1996} we obtain the direction of sensitivity using the
Fr\'echet derivative of a mapping between two normed spaces.
Throughout the rest of the paper we denote the sampling density and base prior by $f(x;\mu)$ and $\pi_{0}(\mu;\theta)$,
respectively, and $x=(x_1,\cdots,x_n)$ represents the vector of observations.
\begin{lemma}\label{perturbed_post}
Under the prior perturbation (\ref{local_perturb2}), the perturbed posterior model is
\begin{eqnarray}
\pi_{p}(\mu,\lambda|x;\theta) = \frac{\pi_{p}^{0}(\mu|x,\theta)}{\xi(\lambda,\theta)} \left\{1+\sum\nolimits_{j=2}^{k}{\lambda_j \, q_j(\mu,\theta)}\right\},\hspace{.5cm} \lambda \in \Lambda_{\theta} \label{post_perturb2}
\end{eqnarray}
with $\xi(\lambda,\theta)=1+\sum_{j=2}^{k}{\lambda_j \, E^0_p[q_j(\mu,\theta)]}>0$,
where $\pi_{p}^{0}(\mu|x,\theta)$ and $E^0_p(\cdot|x)$ are the posterior density and posterior mean of the base model.
\end{lemma}
The following lemma characterizes the $l^{th}$ moment of the perturbed posterior model. Note that,
throughout the rest of the paper, for simplicity of exposition, we  suppress the explicit dependence of $\xi$, $q_j$, $\pi_{p}^{0}$ and
$\pi_{p}$ on $\theta$. 
\begin{lemma}\label{perturbed_moments}       
The moments of the perturbed posterior distribution are given by
\begin{eqnarray}
E_{p}(\mu^l|x,\lambda) = \frac{1}{\xi(\lambda)} \left\{E_{p}^{0}(\mu^l) + \sum\nolimits_{j=2}^{k}{\lambda_j\,A_j^l (x)}\right\},\quad\quad \lambda\in \Lambda_{\theta}\label{post_means}.
\end{eqnarray}
where $A_j^l (x)=E^0_p(\mu^l\,q_j(\mu)|x)$.
\end{lemma}
To  quantify the magnitude of perturbation we exploit the  size function as defined in \citet*{Gustafson1996}, i.e.,
the $L^p$ norm of the ratio $\frac{u^*}{\pi_0}$, for $p<\infty$, with respect to the induced measure by $\pi_0$.
Accordingly, the size function for $u(\cdot)$ is
$$\text{size}(u)=\left[E_{\pi_0}\left(\left|\sum\nolimits_{j=2}^{k}{\lambda_j \, q_j(\mu)}\right|\right)^p\right]^{\frac{1}{p}},$$
which, (i)  is a finite norm and (ii)  is invariant with respect to change of the dominating measure and also with respect
to any one-to-one transformation on the sample space.
Clearly, $\text{size}(u)$ is finite if the first $k+p$ moments of $\pi_0(\mu,\theta)$ exist.
In addition, property (ii) holds by use of change of variable formula and the fact that for any one-to-one
transformation $m=\nu(\mu)$ we have
$\bar{\pi}_0^{(j)}(m,\theta)/\bar{\pi}_0(m,\theta)=\pi_0^{(j)}(\mu,\theta)/\pi_0(\mu,\theta).$

For a mapping $T:\mathcal{U}\rightarrow \mathcal{V}$, where $\mathcal{U}$ and $\mathcal{V}$ are, respectively, the
perturbations space normed with $\text{size}(\cdot)$, and the space of posterior expectations normed with absolute value, the
Fr\'echet derivative at $u_0\in\mathcal{U}$ is defined by the linear functional $\dot{T}(u_0):\mathcal{U}\rightarrow \mathcal{V}$ satisfying 
$$||T(u_0+u)-T(u_0)-\dot{T}(u_0)u||_{\mathcal{V}}=o(||u||_{\mathcal{U}}),$$
in which $\dot{T}(u_0)u$ is the rate of change of $T$ at $u_0$ in direction $u$.
Let $Cov_p^{0}(\cdot,\cdot)$ be the posterior covariance with respect to the base model, 
Theorem \ref{varphi_form} expresses $\dot{T}(u_0)u$ as a linear function of $\lambda$, at $u_0=0$ which corresponds
to the base prior model.
\begin{theorem}\label{varphi_form}
$\dot{T}(0)u$ is a linear function of $\lambda$ as
 \begin{eqnarray}
 \varphi(\lambda)= \sum\nolimits_{j=2}^{k}{\lambda_j\, Cov_p^{0}\left(\mu,q_j(\mu)\right)}, \hspace{1cm} \lambda\in \Lambda_{\theta}\label{var_phi}.
 \end{eqnarray}
\end{theorem}

\section{Global sensitivity}\label{Global sensitivity}
Here we use two commonly applied  measures of sensitivity --  the posterior mean difference and Kullback-Leibler
divergence function --  for assessing the global influence of prior perturbation on posterior mean and
prediction, respectively. 
The following theorem characterizes the difference between the  posterior mean of the base and
perturbed models as a function of $\lambda$.
\begin{theorem}\label{Psi_theorem}
Let $\Psi(\lambda) = E_p(\mu|x,\lambda)-E_p^0(\mu|x)$ represents the difference between the posterior expectations,
then  
\begin{eqnarray}
\Psi(\lambda)=\frac{1}{\xi(\lambda)}\,\varphi(\lambda),  \hspace{1cm}\lambda\in \Lambda_\theta \label{glob_diff}.
\end{eqnarray}
\end{theorem}
The function in (\ref{glob_diff}) behaves in a  intuitively natural way,  for as $\lambda\rightarrow 0$ we have $\xi(\lambda)\rightarrow 1$,
and consequently $\Psi(\lambda)$ behaves locally in a similar way  to $\varphi(\lambda)$.

To assess the influence  of the prior perturbation on prediction, we quantify the change in the  divergence in the posterior
predictive distribution. 

As a illustrative  example, suppose the sampling distribution and the base prior model are respectively
$\mathcal{N}(\mu,\sigma^2)$ and $\mathcal{N}(\theta,\sigma_0^2)$. The posterior predictive distribution for the base model
is $\mathcal{N}(\mu_{\pi},\sigma^2_{\pi}+\sigma^2)$,
where 
$$ \mu_{\pi}= \frac{\theta\sigma^2+n \sigma_0^2\bar{x}}{n\sigma_0^2+\sigma^2}\hspace{.5cm},\hspace{.5cm} \sigma^2_{\pi}=\frac{\sigma^2 \sigma_0^2}{n\sigma_0^2+\sigma^2}.$$
\begin{lemma}\label{posterior_predict}
The posterior predictive distribution for the perturbed model is 
\begin{eqnarray}
g_{p}(y)
=\frac{1}{\xi(\lambda)}\left\{g_{p}^{0}(y)+  \Gamma \sum\nolimits_{j=2}^{k}{   \lambda_j \, E^{\star}[q_j(\mu)]}\right\}\label{g_new_final}
\end{eqnarray}
in which, $g_{p}^{0}(y)$ is the posterior predictive density for the base model, $\Gamma$ is a function of $(y,x,n,\theta_0,\sigma_0^{2},\sigma^2)$
and $E^{\star}(\cdot)$ is expectation with respect to a normal distribution.
\end{lemma}
For probability measures $P_0$ and $P_1$ with the same support space, $S$, and densities $g^0_{p}(\cdot)$ and $g_{p}(\cdot)$,
respectively, Kullback-Leibler divergence functional is defined by, 
\begin{eqnarray}
D_{KL}(P_0,P_1)=\int_S \log\left[g^0_{p}(y)/g_{p}(y)\right]g^0_{p}(y) \, dy \label{K_L derivation0}
\end{eqnarray}
which satisfies the following conditions (see \cite{Amari1990}), 
\begin{enumerate}
 \item $D_{KL}(P_0,P_1)\geq 0$, with equality if and only if $P_0\equiv P_1$.
 \item $D_{KL}(P_0,P_1)$ is invariant under any transformation of the sample space.
\end{enumerate}

\begin{theorem}\label{Diverg_thorem}
Kullback-Leibler divergence between $g^0_{p}(\cdot)$ and $g_{p}(\cdot)$ as a function of $\lambda$ is 
\begin{eqnarray}
D_{KL}(\lambda)
&=& \int_S \log\left[g^0_{p}(y)\right] \,g^0_{p}(y) \,dy + \log[\xi(\lambda)]\nonumber\\
&&- \int_S \log\left(g_{p}^{0}(y)+  \Gamma \sum\nolimits_{j=2}^{k}{   \lambda_j  E^{\star}[q_j(\mu)]}\right)g^0_{p}(y)  dy ,  \hspace{1cm}\lambda\in \Lambda_\theta\label{D_KL}
\end{eqnarray} 
\end{theorem}

\section{Estimating $\lambda$}\label{estimating lambda}
To obtain the values of $\lambda$ which find the most sensitive local and global perturbations,  as described in Section \ref{Introduction}, we apply an
optimization approach to the functions (\ref{var_phi}), (\ref{glob_diff}) and (\ref{D_KL}).
$\varphi(\lambda)$ is a linear function of $\lambda$ on the space $\Lambda_{\theta}$ which presents the directional
derivative of the mapping $T$ at $\lambda=0$. Thus, for obtaining the maximum direction of sensitivity,
called the worst local sensitivity direction in \cite{Gustafson1996}, we need to maximize $\varphi(\lambda)$ over
all the possible directions at $\lambda=0$ restricted by the boundary of $\Lambda_{\theta}$. However, $\Psi(\lambda)$
and $D_{KL}(\lambda)$ are smooth objective functions on the convex space $\Lambda_\theta$, for which we propose a suitable
gradient based constraint optimization method that exploits the geometry of the parameter space. 
By Definition \ref{loc_mix}, for a fixed known $\theta$, the space $\Lambda_{\theta}$ is a non-empty convex subspace
in $R^{k-1}$ with its boundary obtained by the following infinite set of hyperplanes
$$\mathcal{H}=\left\{\lambda\,\, \Bigl|\,\,1 + \sum\nolimits_{j=2}^{k}{\lambda_j \, q_j(\mu)}=0\,\, ;\,\, \mu\in R\right\}.$$
Specifically, for the normal example with order $k=4$, $\mathcal{H}$ is the infinite set of planes of the form 
\begin{eqnarray}
P_{\lambda}(z)=\left(z^{2}-\frac{1}{\sigma_0^2}\right)\lambda_2+\left(z^{3} - \frac{3z}{\sigma_0^2}\right)\lambda_3+\left(z^{4} - \frac{6z^{2}}{\sigma_0^2}+\frac{3}{\sigma_0^4}\right)\lambda_4  +1.\label{boundary plane} 
\end{eqnarray}
where $z=\frac{\mu-\theta}{\sigma_0^2}$.
Lemma \ref{Manifold} describes the boundary of $\Lambda_{\theta}$ as a smooth manifold.
\begin{lemma}\label{Manifold}
The boundary of $\Lambda_{\theta}$ is a manifold (smooth surface) embedded in $R^3$ Euclidean space.
\end{lemma}
In addition, the interior of $\Lambda_\theta$, which guarantees positivity of $\pi(\mu,\lambda;\theta)$
for all $\mu \in R$, can be characterized by the necessary and sufficient positivity conditions of
general polynomials of degree four.  The corresponding polynomial to Equation (\ref{boundary plane}) is a quartic
with highest degree coefficient $\lambda_4$; hence, the necessary positivity condition is $\lambda_4>0$. Also the comprehensive 
necessary and sufficient conditions are given in \citet*{Barnard1936} and \citet*{Bandy1966}.  
Throughout the rest of the paper we let $k=4$, as it gives a perturbation space which is flexible enough for our analysis
and it has been illustrated in \cite{Marriott2002}, through examples, that simply increasing the order of local mixture
models does not significantly increase flexibility. However, all the results and algorithms can be generalized to higher
dimensions with possible generalization of the positivity conditions on polynomials with higher degrees.

\begin{lemma}\label{direc_max}
 $\varphi(\lambda)$ attains its maximum value at the gradient direction $\nabla \varphi$ if it is feasible; otherwise,
the maximum direction is the direction of the orthogonal projection of $\nabla \varphi$ onto the boundary plane
corresponding to $\lambda_4=0$.
\end{lemma}

$D_{KL}(\lambda)$ and $\Psi(\lambda)$ are smooth functions which  can achieve their maximum either in the interior
or on the smooth boundary of $\Lambda_{\theta}$. Therefore, optimization shall be implemented in two steps: searching
the interior using regular Newton-Raphson algorithm, and then searching the boundary using a generalized form
of Newton-Raphson algorithm on smooth manifolds, see \cite{Shub1986} and also  \cite{Maroufy}.

\section{Examples}\label{Examples}
We consider three examples, where  the first two study local and global sensitivity in the normal conjugate model
using the optimization approaches developed earlier to address the questions in Section \ref{Introduction}.
In the last example, we address  sensitivity analysis in finite mixture models with independent conjugate
prior models for all parameters of interest. Rather than  using  an optimization approach, for this example a Markov
Chain Monte Carlo method is used and sensitivity of the posterior distribution of each parameter is assessed.
For demonstrating the effect of the perturbation obtained in each example we  compare the posterior
distributions before and after perturbation and also use the relative difference between the Bayes estimates defined by 

$$d=\frac{|E_p^0(\mu)-E_p^{\hat{\lambda}}(\mu)|}{std_p^0(\mu)}$$
in which $E_p^0(\mu)$ and $E_p^{\hat{\lambda}}(\mu)$ are the Bayes estimates with respect to the base
and perturbed models, respectively, and $std_p^0(\mu)$ is the posterior standard deviation under the
base model. Since $\Psi(\lambda)$ also allows negative values, care must be taken as we may need to
minimize this function instead of maximizing it for achieving the maximum discrepancy between the posterior
distributions.

\begin{example}[Normal conjugate]\label{Example1}
A sample of size $n=15$ is taken from $\mathcal{N}(1,1)$, and the base prior is $\mathcal{N}(2,1)$. The estimate
$\hat{\lambda}_{D}=(1.821,-0.014,0.482)$ and $\hat{\lambda}_{\Psi}=(1.817,-0.009,0.486)$ are obtained
from maximizing $D_{KL}(\lambda)$ and minimizing $\Psi(\lambda)$, respectively. The corresponding relative
discrepancies in Bayes estimate are respectively $d=1.19,1.2$; that is, the resulted biases in estimating
posterior expectation are respectively $119\%$ and $120\%$ of the posterior standard deviation of the base
model. Also, the corresponding posterior distributions are plotted in Figure \ref{Ex1_plot}.
 Considering the fact that we construct the perturbation space as a family of local mixture models which
are close to the base prior model, these maximum global perturbations are obtained by searching over a reasonably
small space of prior distributions which only different from the base prior by their tail behaviour.
Hence, these results imply that although conjugate priors are convenient in applications, they might cause significant
bias in estimation as a result of even plausibly small prior perturbations.

\begin{figure}[!h]  
  \center
  \includegraphics[scale=.13]{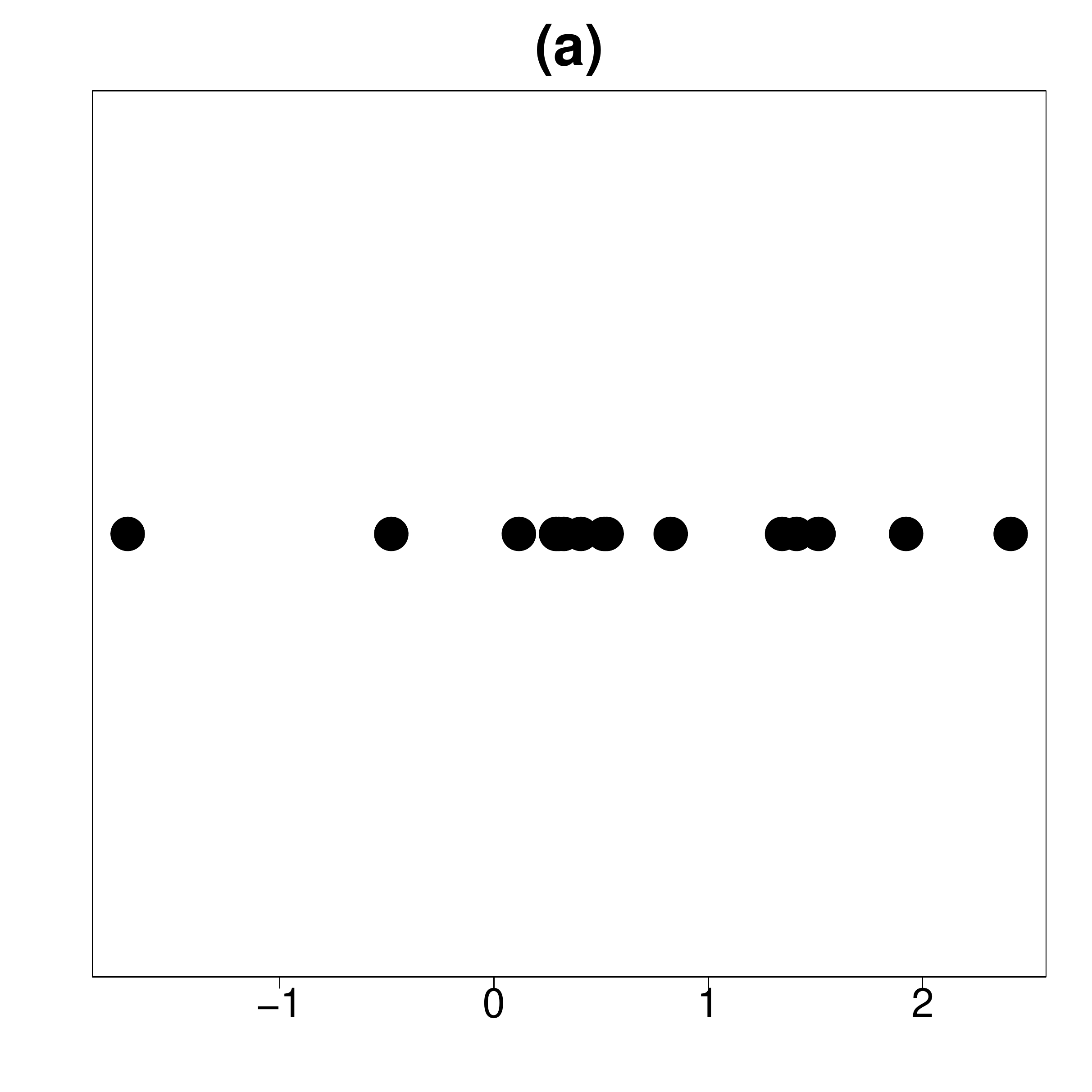}
  \includegraphics[scale=.13]{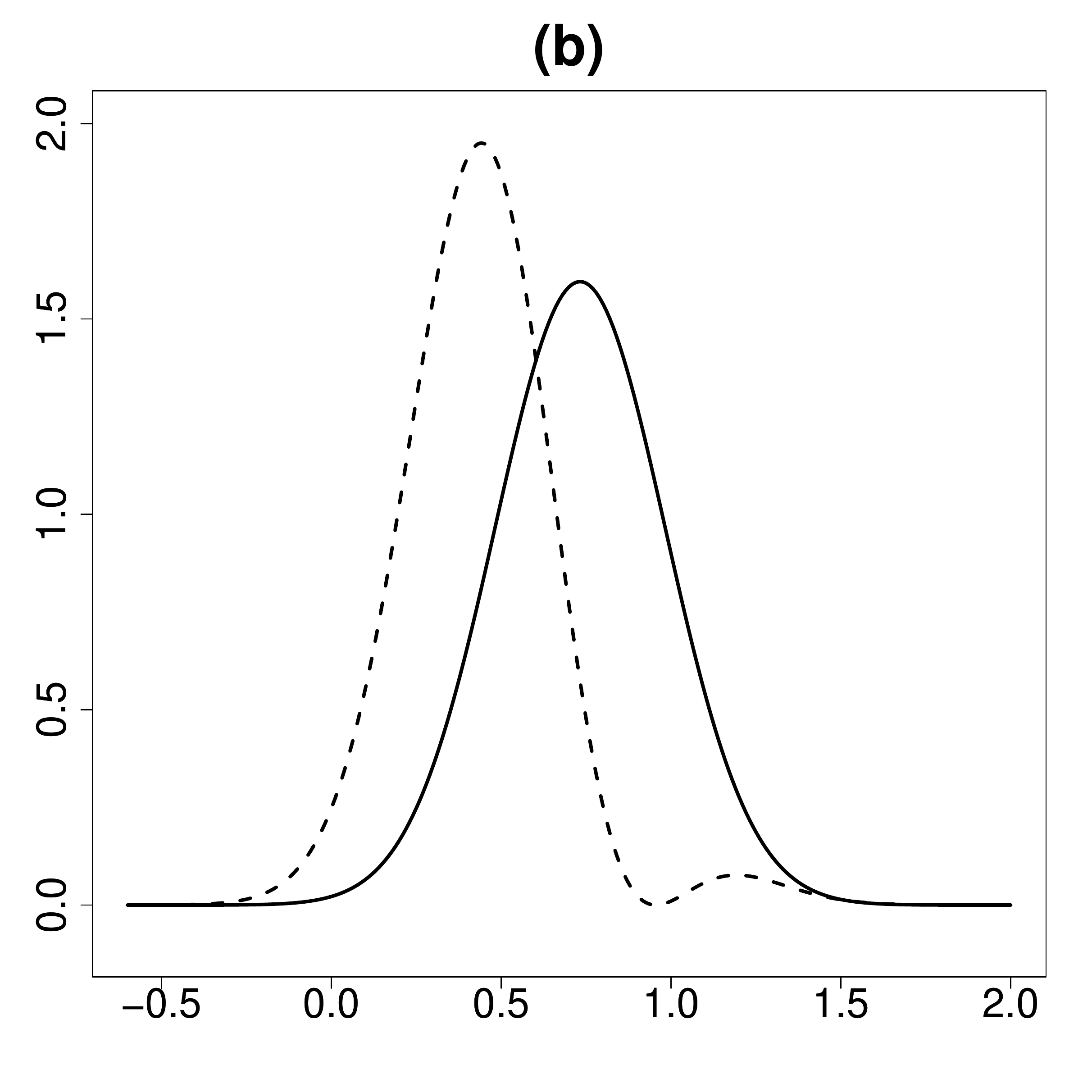}
  \includegraphics[scale=.13]{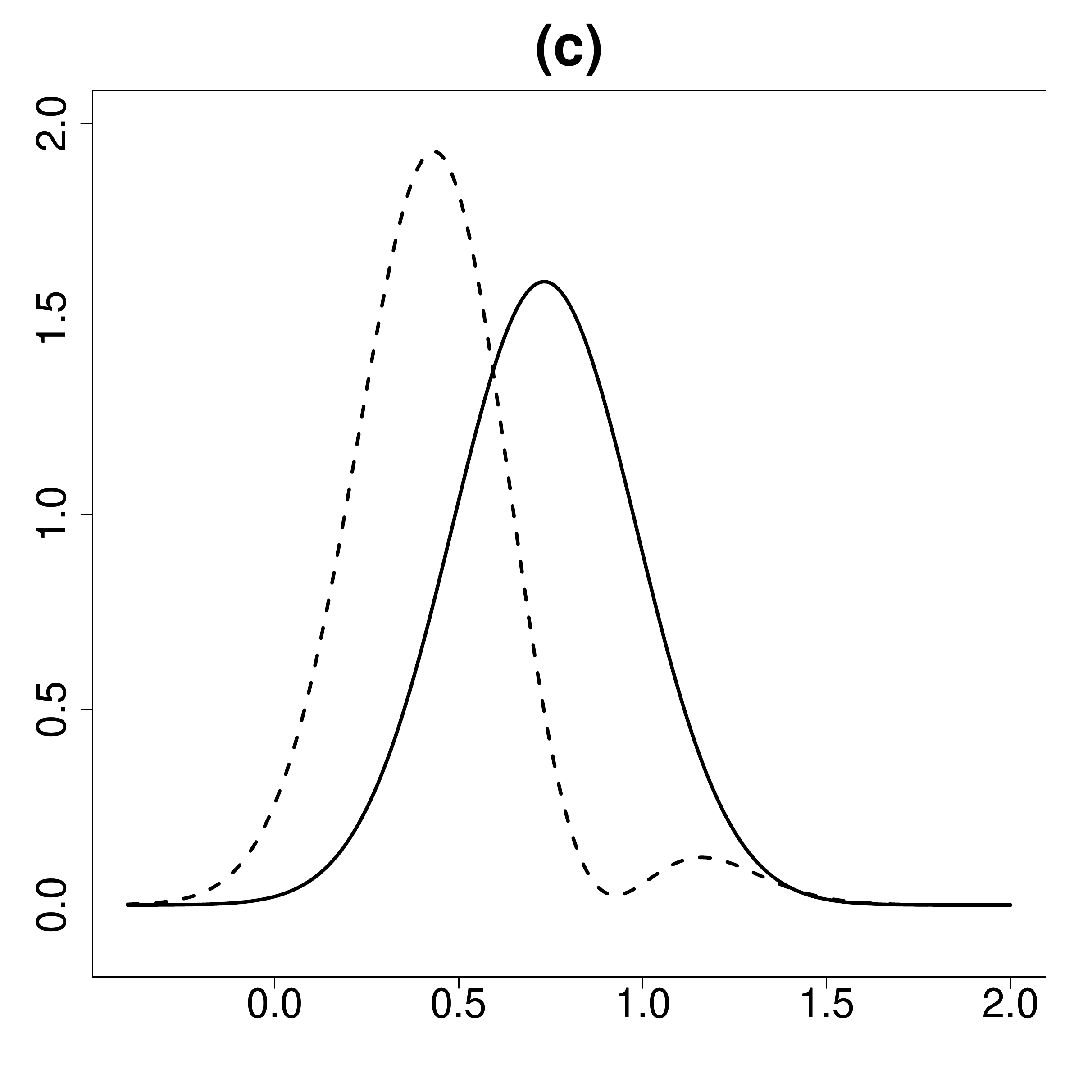}\\
   \caption{ (a) sample, (b) and (c) posterior densities of based models and perturbed model
   (dashed) corresponding to $\hat{\lambda}_{D}$ and $\hat{\lambda}_{\Psi}$ respectively.   }\label{Ex1_plot}
\end{figure}

For local analysis, we obtained the unit  vector $\hat{\lambda}_\varphi$ which maximizes the directional
derivative $\varphi(\lambda)$. Figure \ref{Ex1_local} presents the posterior density displacement 
by perturbation parameter $\lambda_\alpha= \alpha \hat{\lambda}_\varphi$ for different values of $\alpha>0$, as well as
the boundary point $\lambda_b$ in direction of $\hat{\lambda}_\varphi$. The corresponding relative differences
in posterior expectation are $d=0.1,0.16,0.25,0.38,0.49,0.56$. Hence, additional to obtaining the worst 
direction, these values suggest that how far one can perturb the base prior along the worst direction so that
relative discrepancy in posterior mean estimation is less than, say $50\%$. 

\begin{figure}[!h]  
  \center
  \includegraphics[scale=.2]{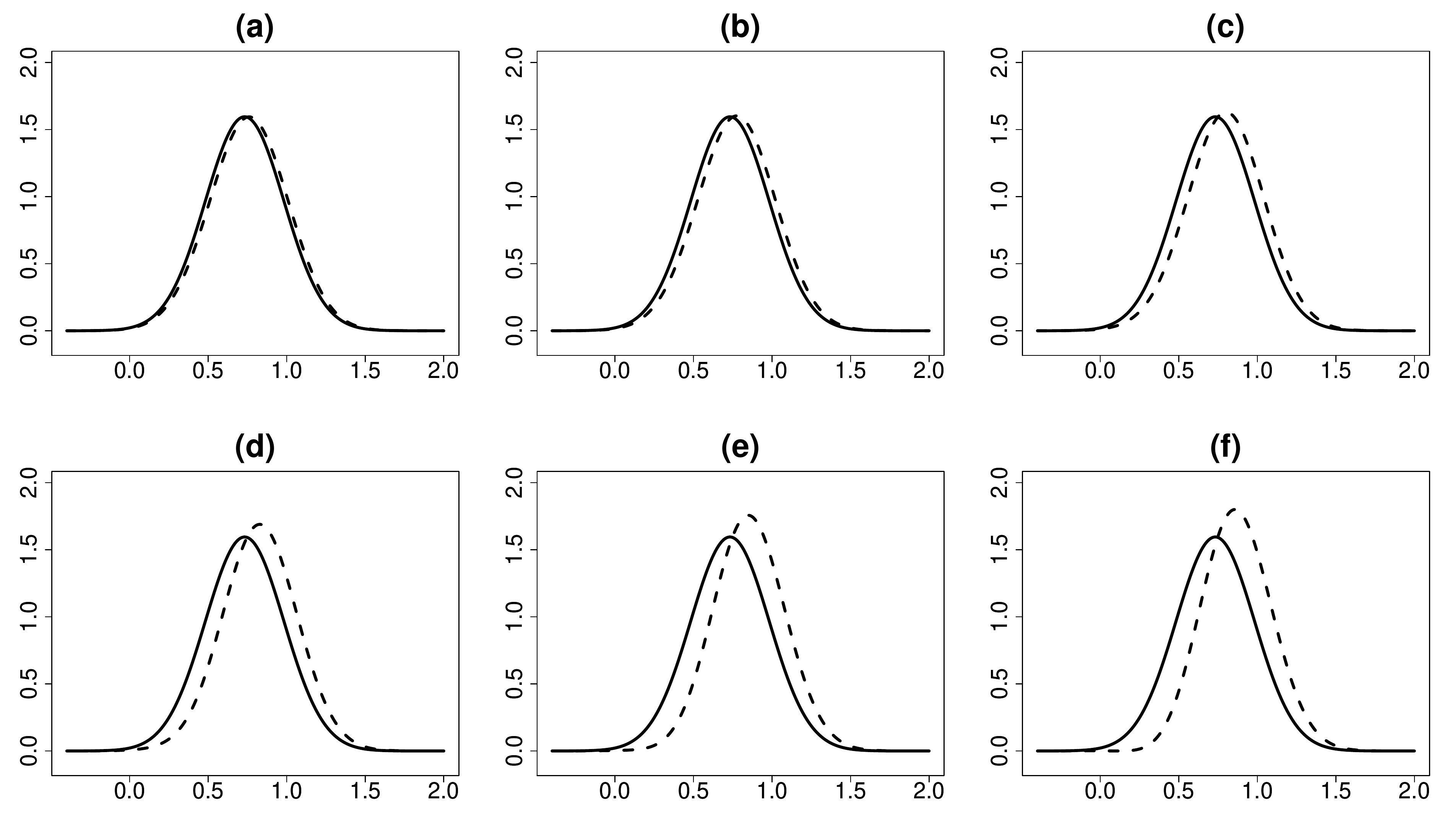}
   \caption{ (a)-(e) posterior densities of based models and perturbed model
   (dashed) corresponding to $\lambda=\alpha \hat{\lambda}_\varphi$ where $\alpha=0.05,0.07,0.1,0.13,0.15$ and (f) for boundary
   point in direction of $\hat{\lambda}_\varphi$.}\label{Ex1_local}
\end{figure}
\end{example}

\begin{example}\label{Example2}
The central moments of the perturbed prior model, in Definition (\ref{perturb_prior}), are linearly related
to the perturbation parameter $\lambda$. Specifically, for the normal model we can check that 
\begin{eqnarray}
\bar{\mu}_{\pi}^{(2)}=\sigma^2+2\lambda_2\,,  \quad\bar{\mu}_{\pi}^{(3)}=6\lambda_3\,,\quad \bar{\mu}_{\pi}^{(4)}=\bar{\mu}_{\pi_0}^{(4)}+12\sigma^2\lambda_2 +24\lambda_4\label{soft1} 
\end{eqnarray}
where $\bar{\mu}_{\pi}^{(j)}$ represents the $jth$ central moment with respect to density $\pi$. Clearly,
$\lambda_2$ modifies variance, $\lambda_3$ adds skewness, and $\lambda_4$ adjusts the tails.

Suppose that elicited prior knowledge tells us that the perturbed prior is required to stay symmetric, then the perturbation space
must be modified by the extra restriction $\lambda_3=0$, which gives zero skewness. Consequently, we should be exploring the restricted space,
say $\Lambda^{0}_\theta$, instead of $\Lambda_\theta$, for the worst direction and maximum global perturbation.
$\Lambda^{0}_\theta$ is a 2-dimensional cross section obtained from intersection of $\Lambda_\theta$ with the
plane defined by $\lambda_3=0$. Hence the boundary properties are preserved.
For the same data in Example \ref{Example1}, sensitivity in the worst direction returns $d=0.1,0.16,0.26,0.42,0.57,0.64$
(Figure \ref{Ex2_local}). Also, minimizing $\Psi(\lambda)|_{\lambda_3=0}$ returns $\hat{\lambda}^0_\Psi=(1.837,0.494)$.

 Two observations can  be made from these results. First, as in Example \ref{Example1}, although we have restricted the
perturbation space further, there are still noticeable discrepancies in posterior densities caused by perturbation along
the worst direction. Second,
the results agree with that in Example \ref{Example1}, where the estimate of $\lambda_3$ does not seem to be significantly
different from zero, and the rest of two parameter estimates are quite close in both examples.

\begin{figure}[!h]  
  \center
  \includegraphics[scale=.2]{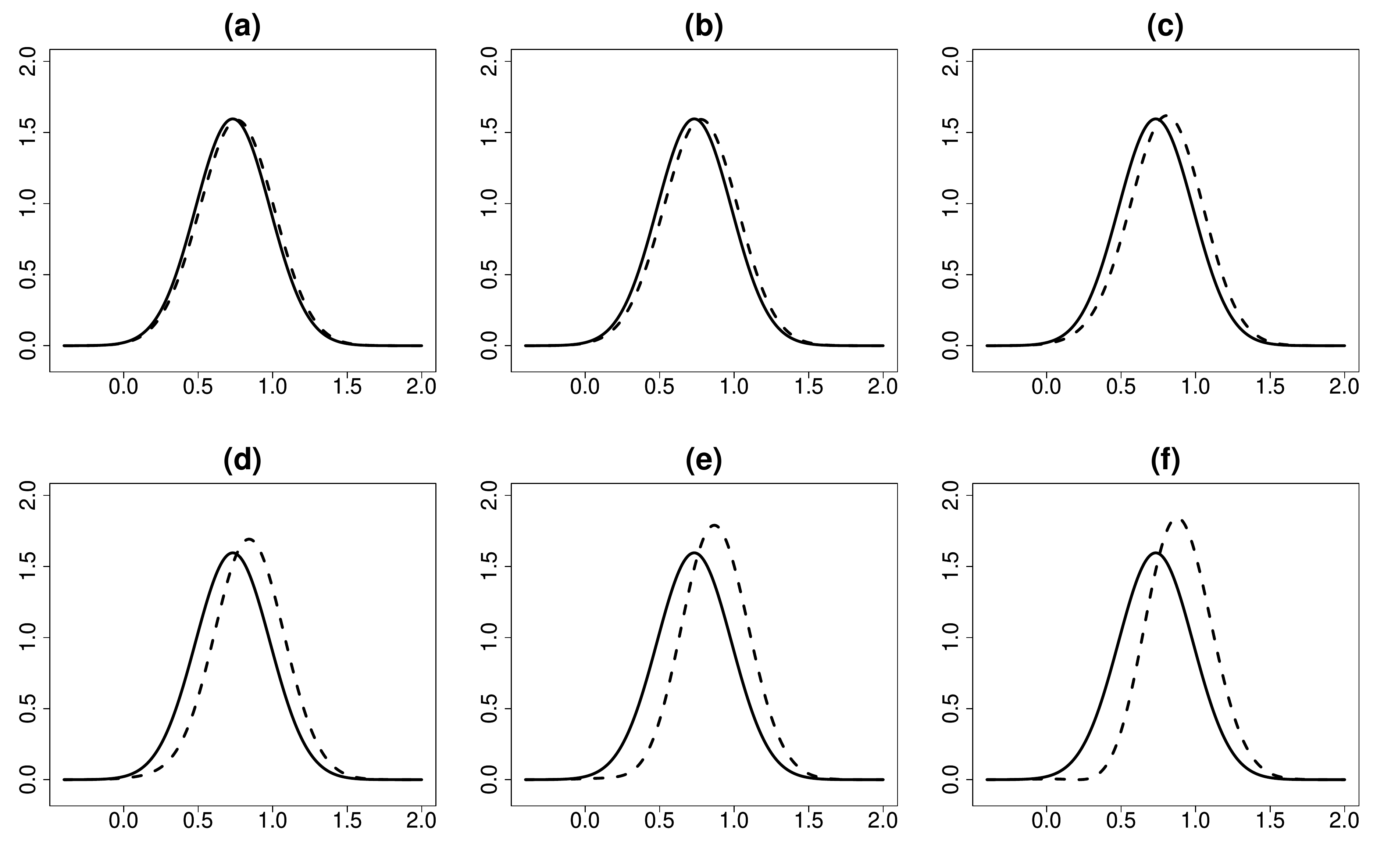}
   \caption{ (a)-(e) posterior densities of based models and perturbed model
   (dashed) corresponding to $\lambda=\alpha \hat{\lambda}_\varphi$ where $\alpha=0.05,0.07,0.1,0.13,0.15$ and (f) for boundary
   point in direction of $\hat{\lambda}_\varphi$.}\label{Ex2_local}
\end{figure}
\end{example}

\begin{example}[Finite Mixture]\label{Example4}
Using a missing value formulation, the likelihood function of the mixture model $\rho\, \mathcal{N}(x;\mu_1,\sigma_1)
+(1-\rho)\, \mathcal{N}(x;\mu_2,\sigma_2)$ can be written as follows
$$L=\prod\nolimits_{j=1}^{2} \rho^{n_j}\prod\nolimits_{i\in A_j} \phi(x_i;\mu_j,\sigma_j),$$
where $A_j=\{i|w_i=j\}$, and $w_i$ is the latent missing variable for $x_i$ such that $p(w_i=1)=\rho$, and $p(w_i=2)=1-\rho$.
The marginal conjugate base prior models are
$\mu_j\sim\mathcal{N}(\theta_j,\sigma_{0j})$, $\sigma_j^{-2}\sim\Gamma(k_j,\tau_j)$, and $\rho\sim Beta(\alpha,\beta)$,
$(j=1,2)$.

In this example the base prior model can be split into five independent components and, correspondingly, five independent
perturbation spaces are naturally defined. Unlike previous examples, where we find the maximum local and global perturbations,
here we explore each marginalized perturbation space by generating perturbation parameters and observing their influence on
the posterior model of parameters of interest. 

Specifically, we use Markov Chain Monte Carlo Gibbs sampling
for estimating the marginal posterior distribution of all parameters of interest, corresponding to the base and perturbed models. Each
perturbation parameter is generated, independently from the rest, through a Metropolis algorithm with a uniform proposal distribution.
Figure \ref{Ex4_plot} shows the histograms of generated sample for an observed data set of  size $n=15$ from $0.4\, \mathcal{N}
(x;-1,1)+0.6\, \mathcal{N}(x;1,1)$, and the hyper-parameters are set to be $\theta_1=-1.5$, $\theta_2=0.5$, $\tau_1=\tau_2=1$, $k=2$
and $\alpha=\beta=1$. Comparing the two histograms for each parameter, the posterior models for $\mu_1$ and $\rho$ show significant
differences between the base and perturbed models. The marginal relative differences are $d=0.49, 0.11,0.40,0.59,0.71$, respectively
for $(\rho,\mu_1,\mu_2,\sigma_1,\sigma_2)$. These differences are not as significant as those in the previous examples
for since they do not correspond to maximum perturbations; instead, they return the average influences over all generated
perturbation parameter values. 

\begin{figure}[!h]  
  \center
  \includegraphics[scale=.16]{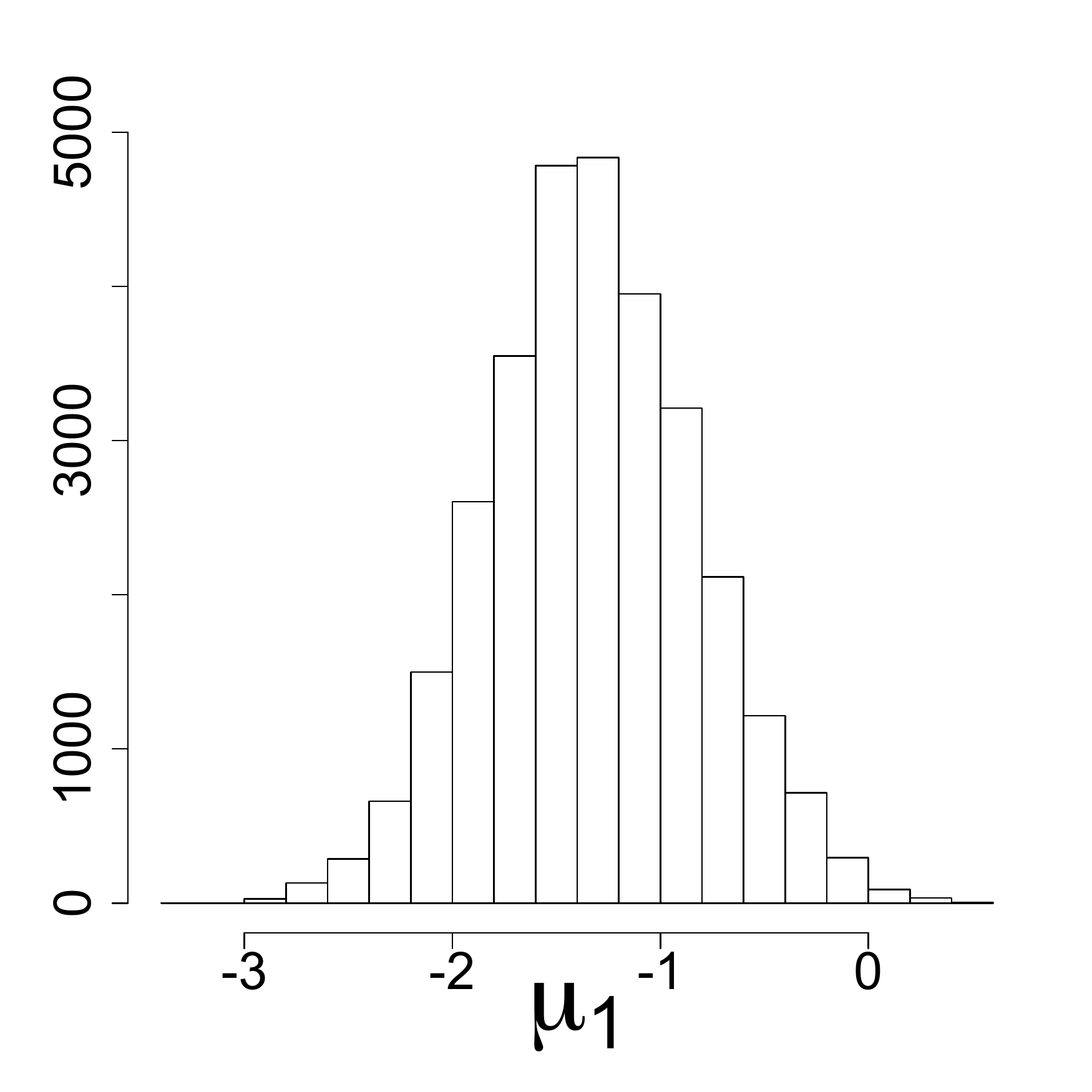}
  \includegraphics[scale=.16]{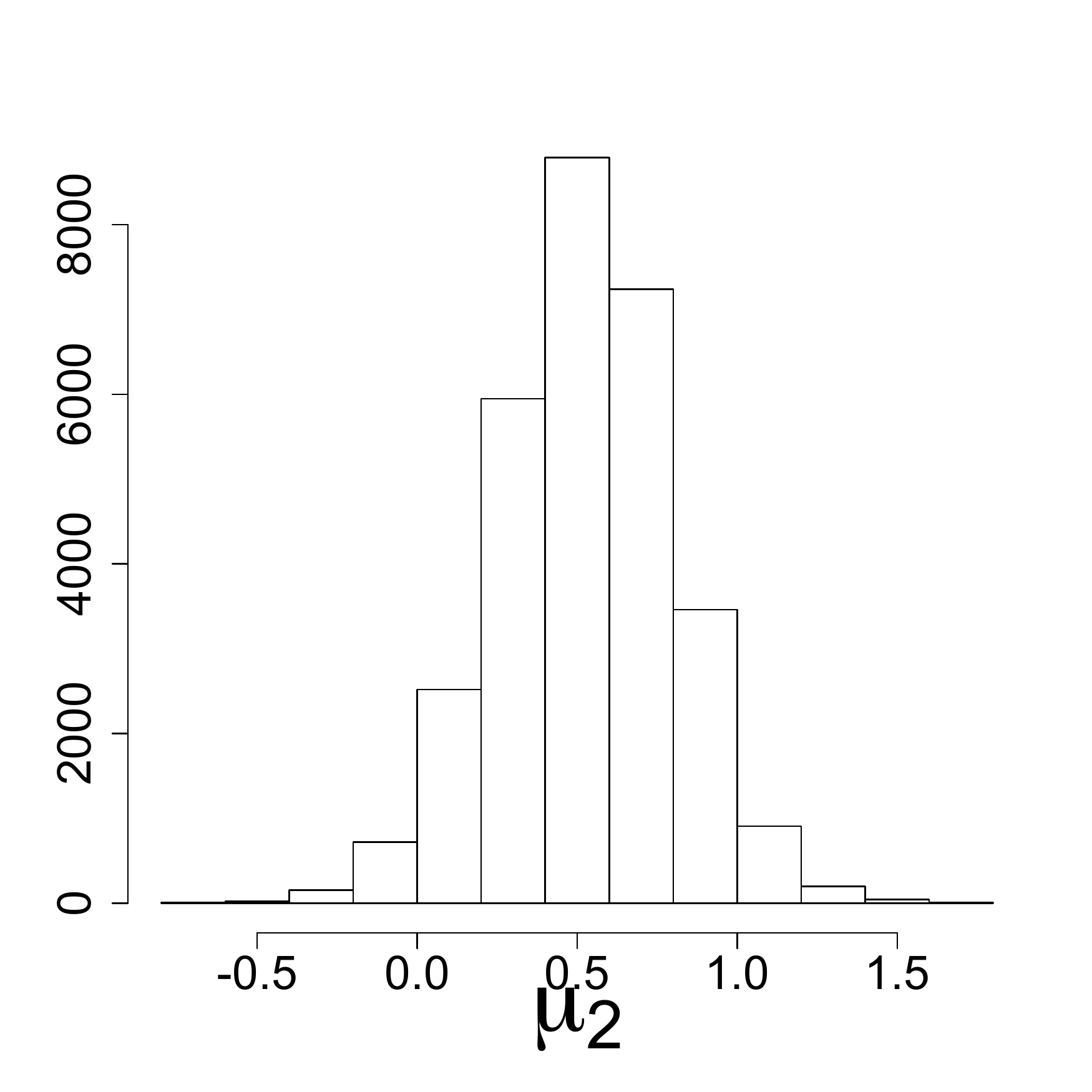}
  \includegraphics[scale=.16]{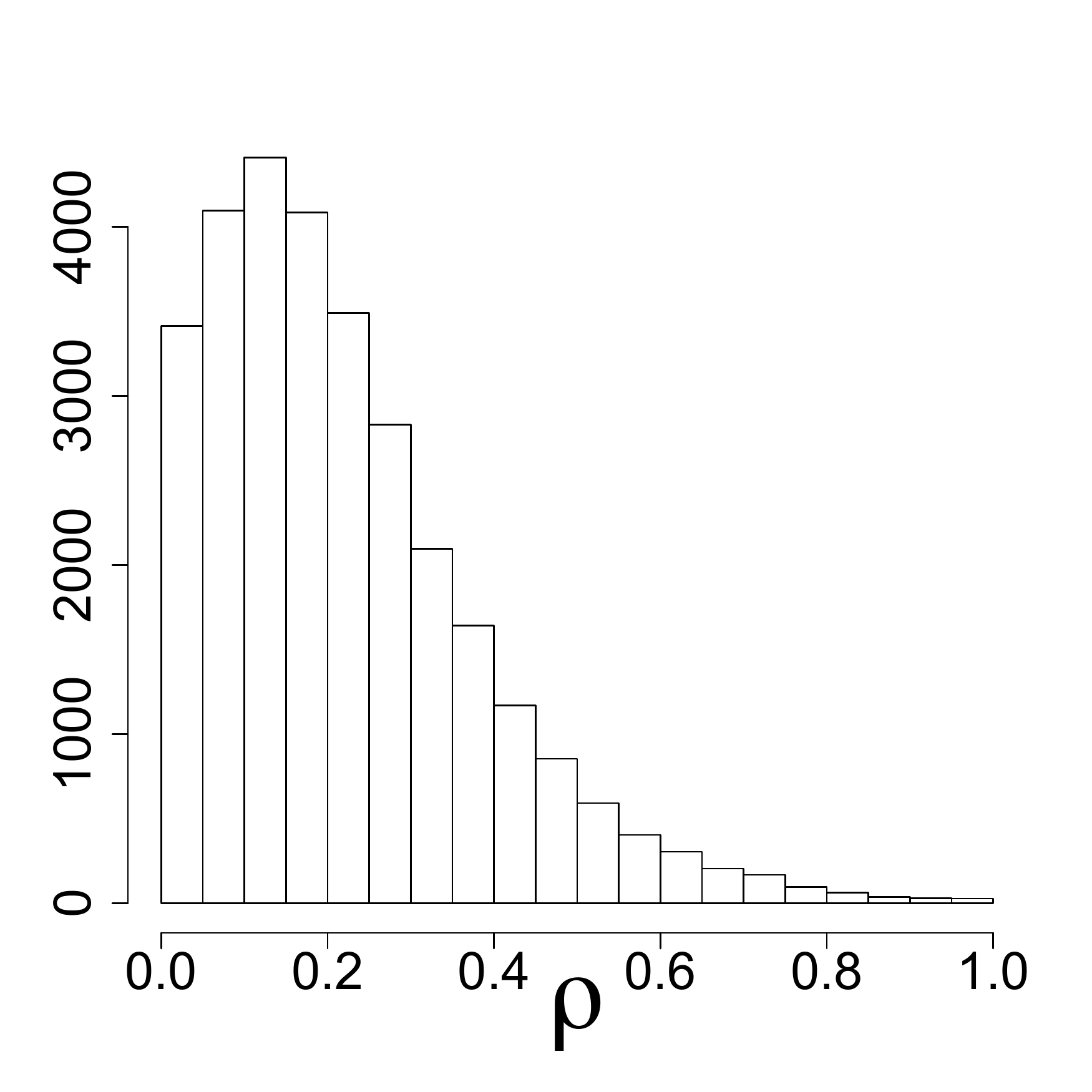}
  \includegraphics[scale=.16]{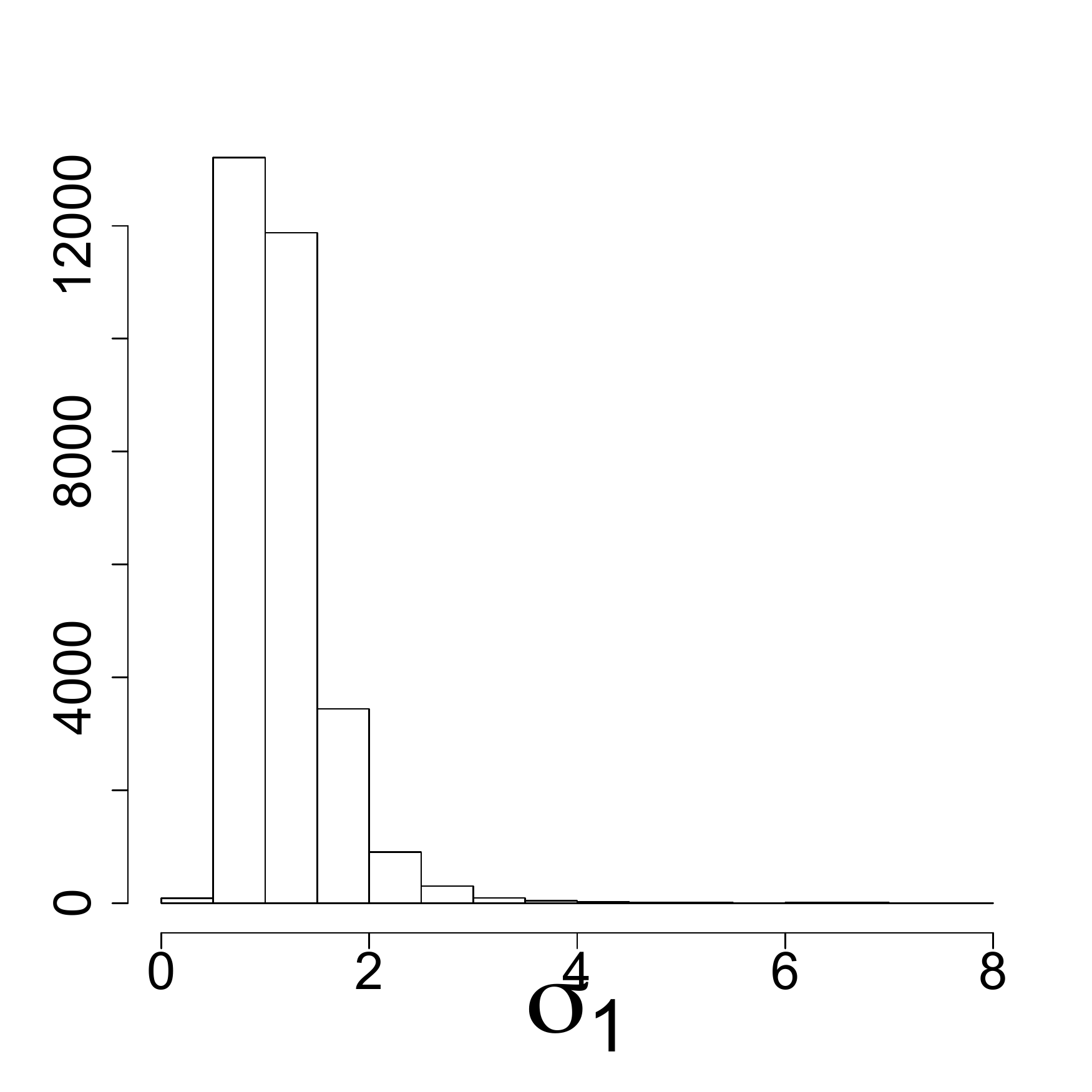}
  \includegraphics[scale=.16]{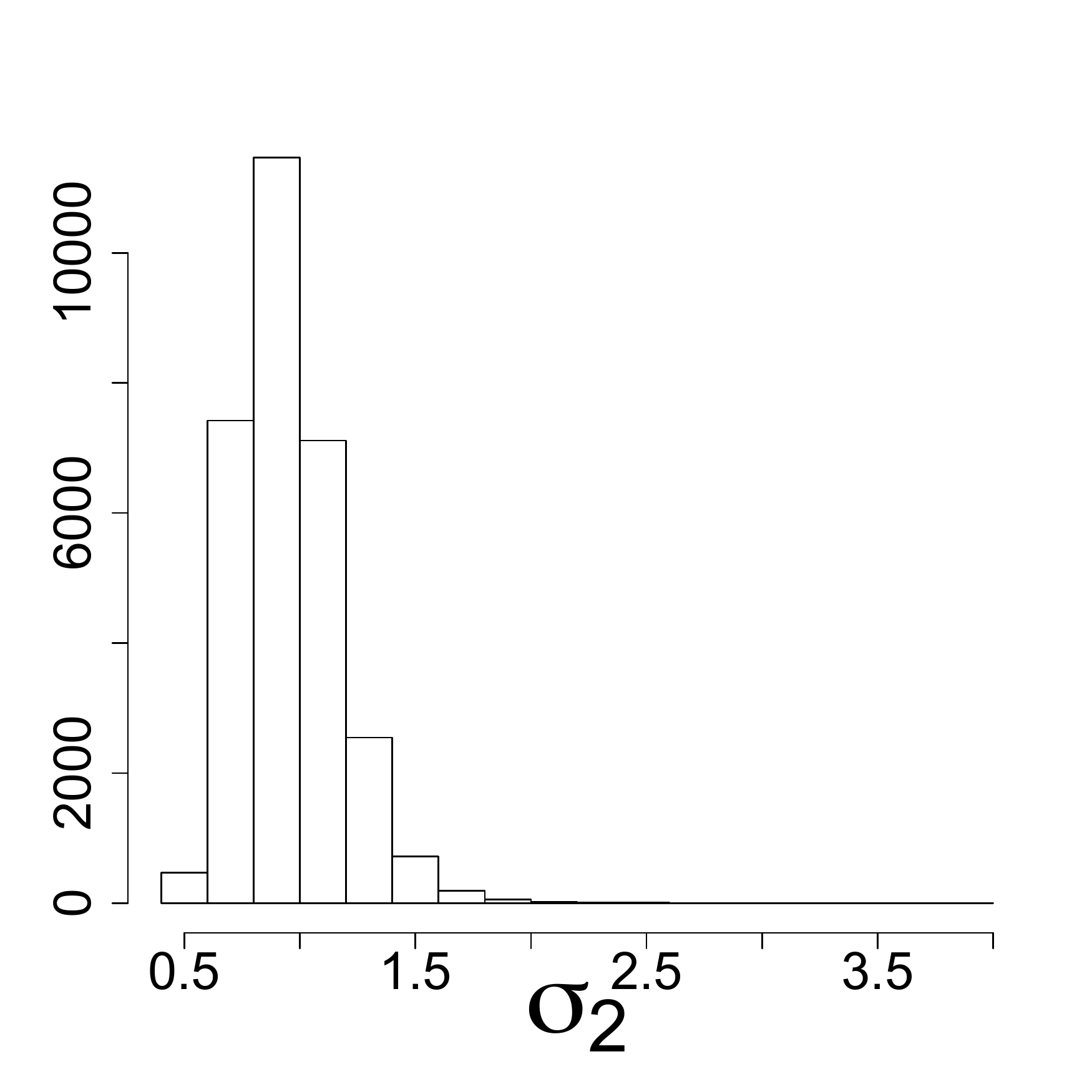}\\
  \includegraphics[scale=.16]{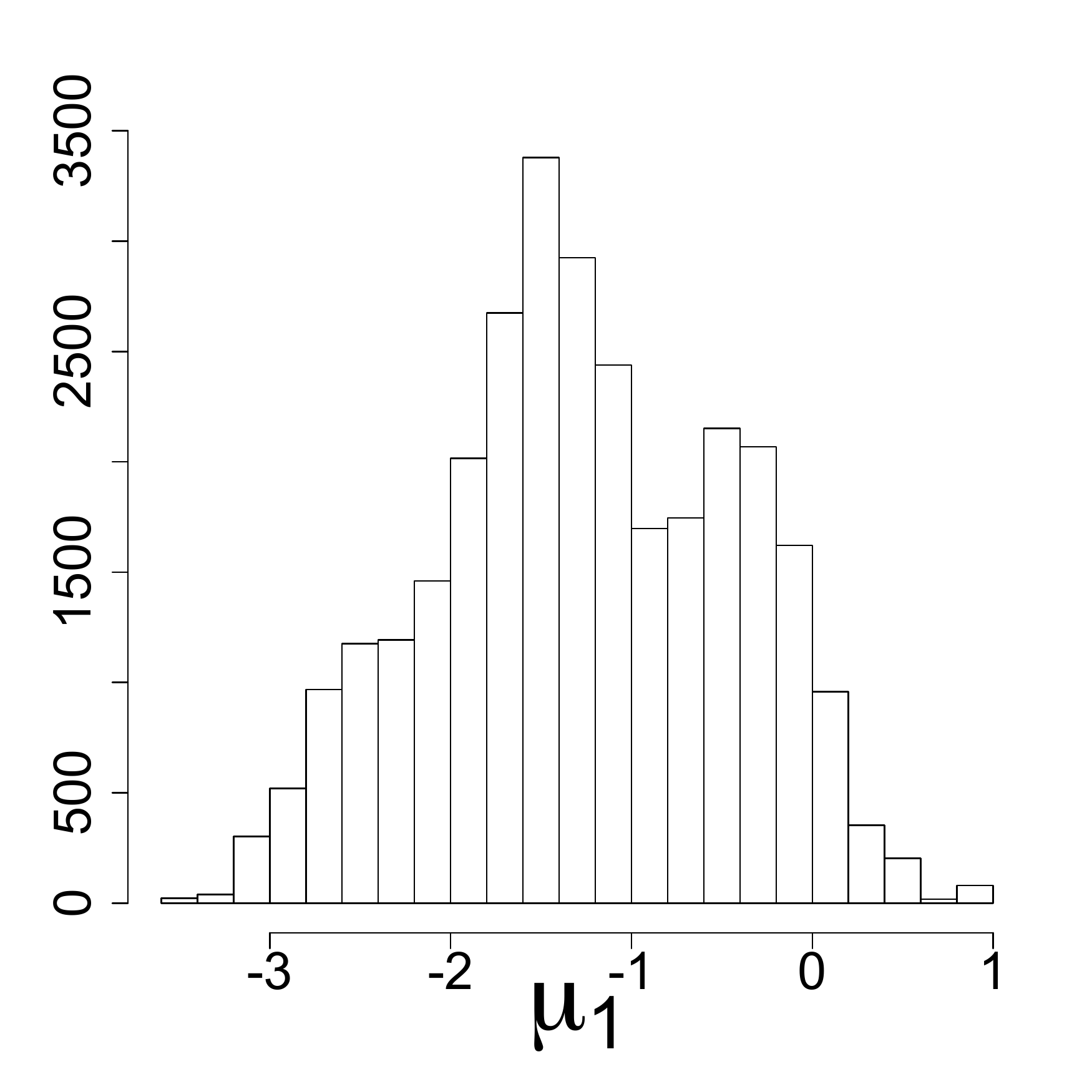}
  \includegraphics[scale=.16]{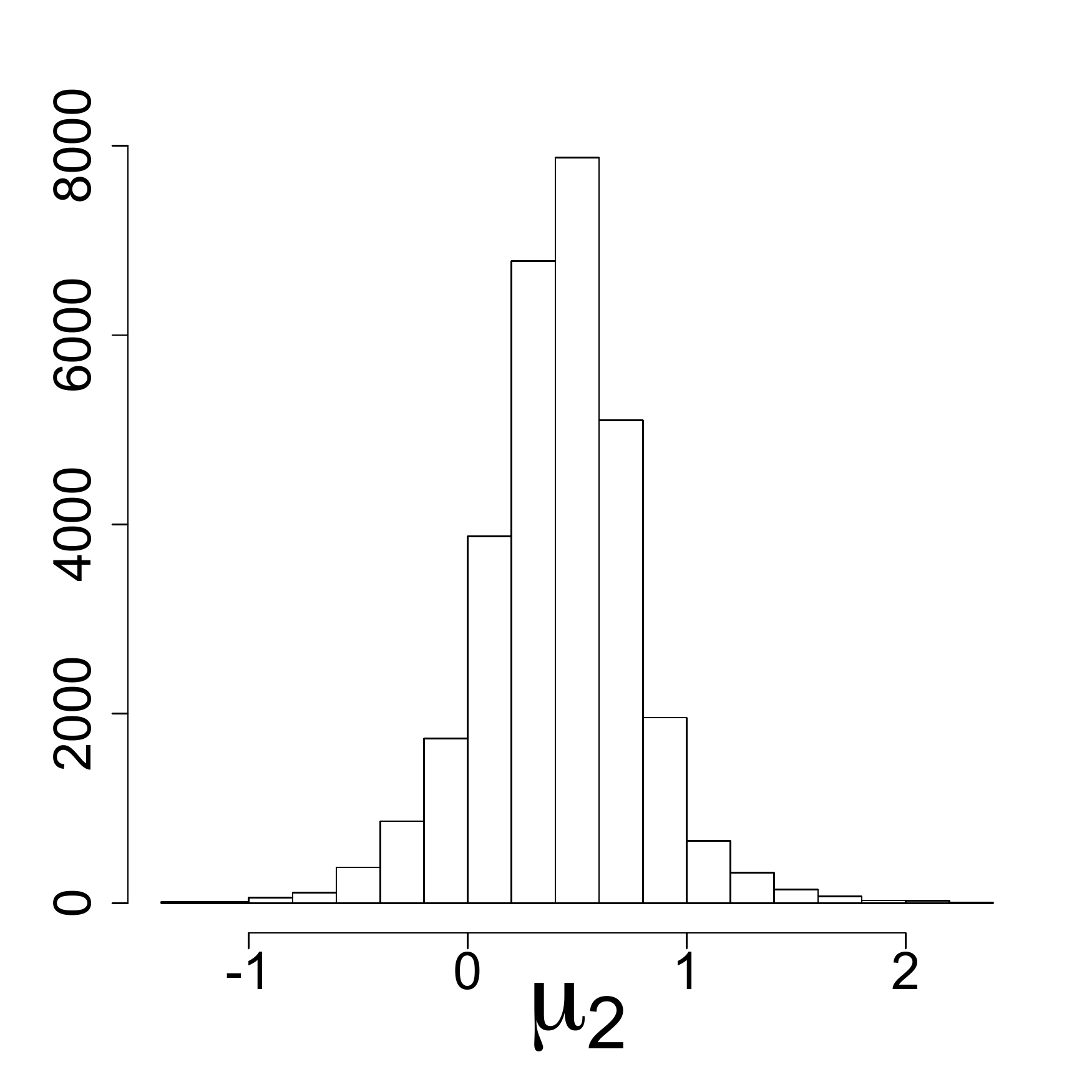}
  \includegraphics[scale=.16]{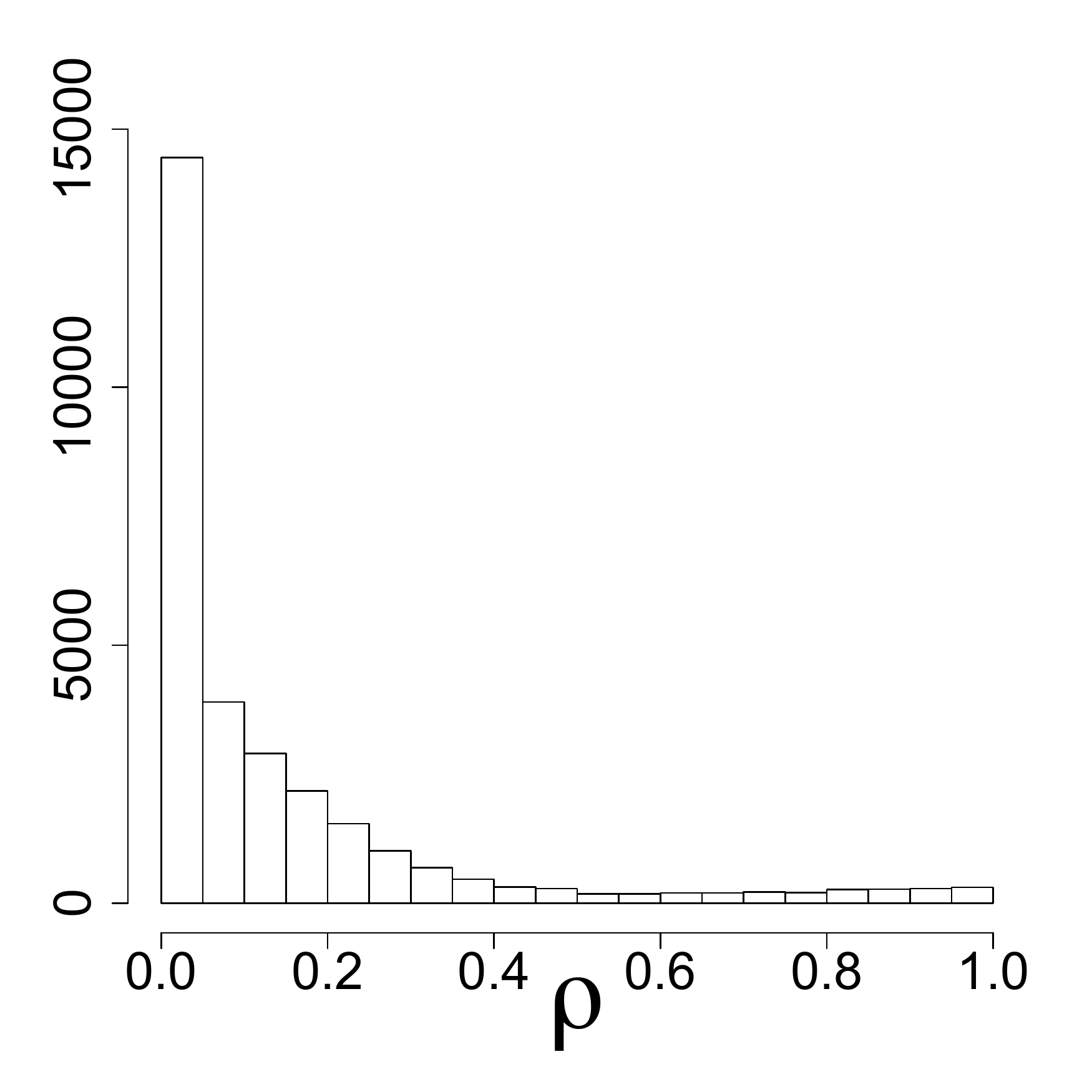}
  \includegraphics[scale=.16]{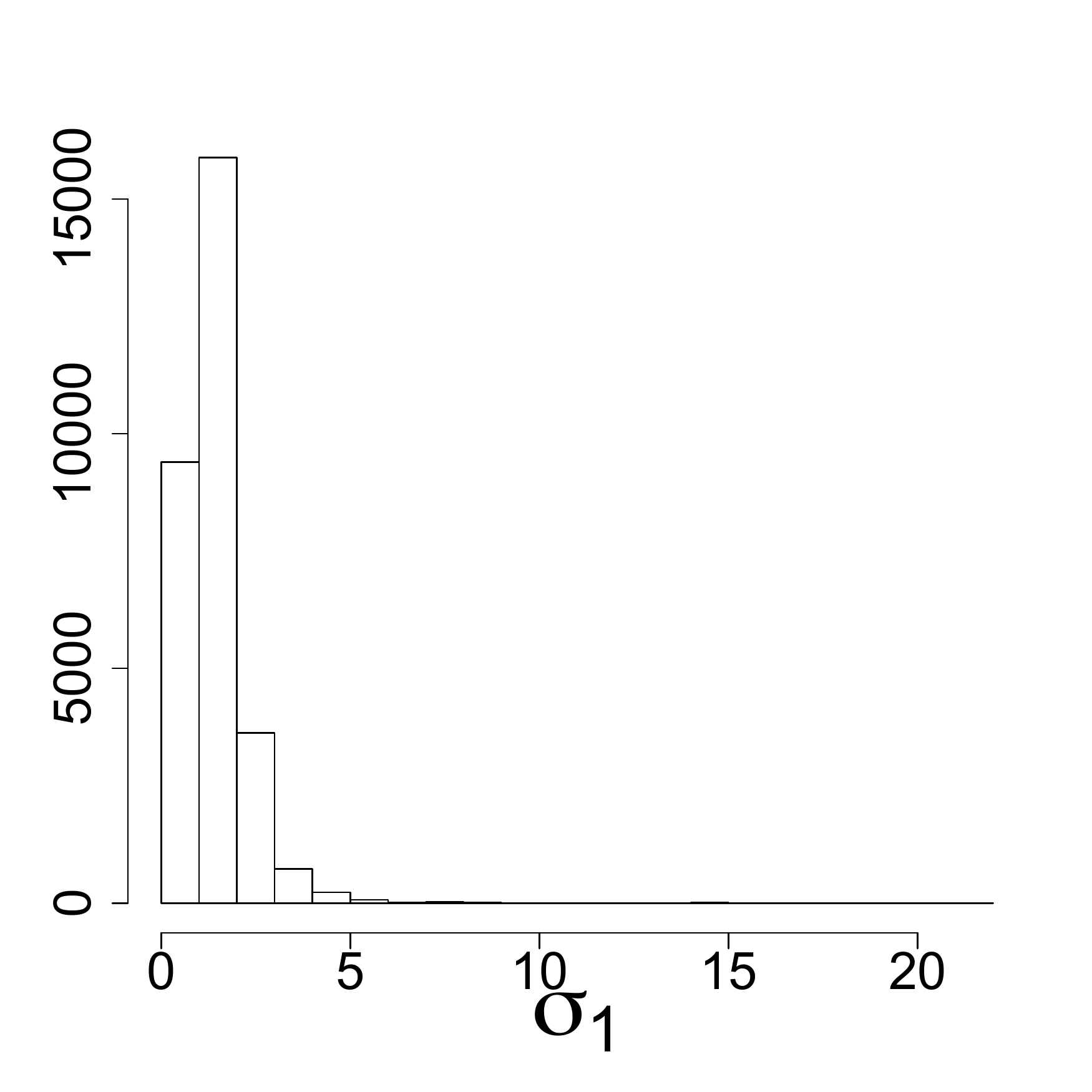}
  \includegraphics[scale=.16]{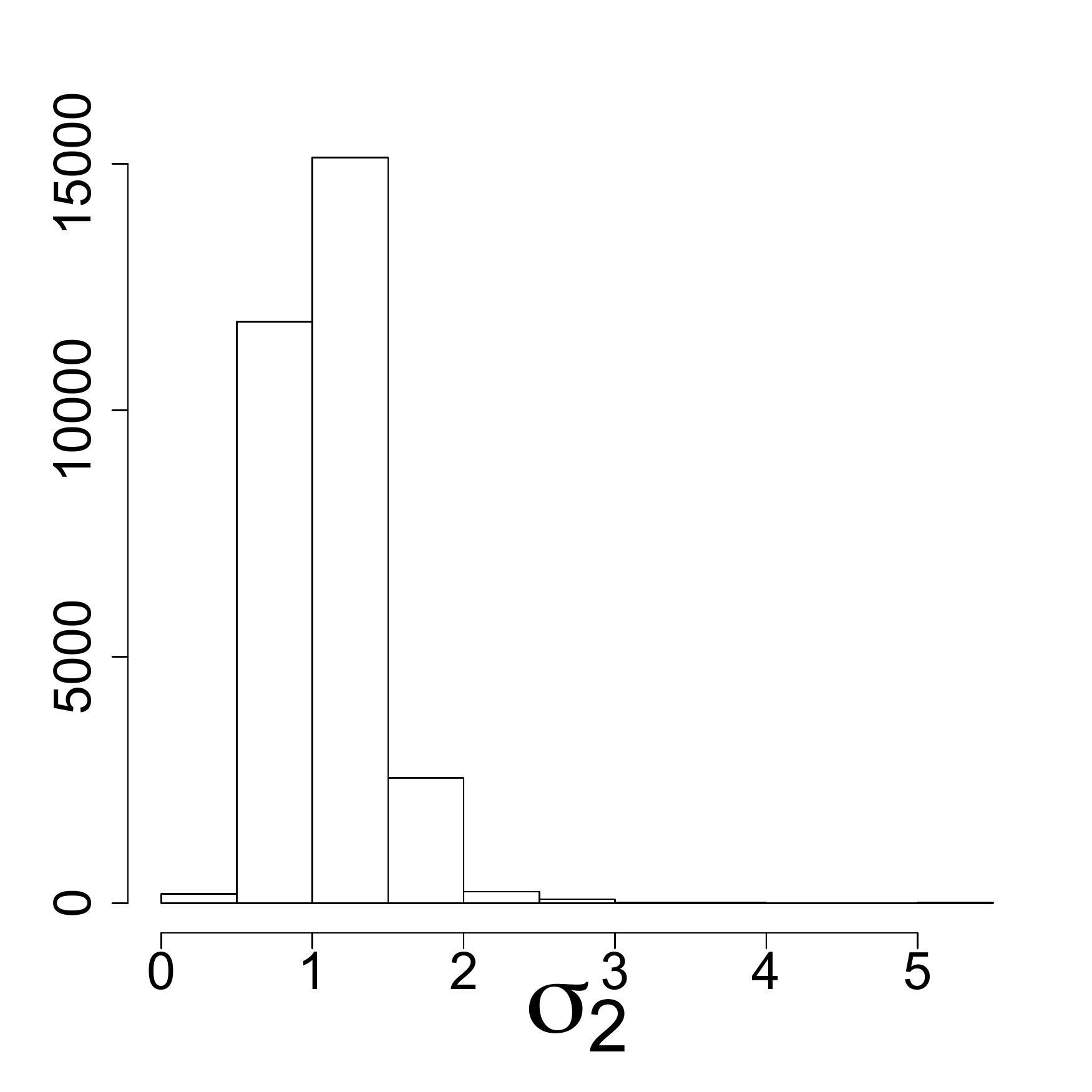}
   \caption{First row: estimates from the base model; second row: estimates form the perturbed model}\label{Ex4_plot}
\end{figure}

\end{example}

The examples of this paper have explored the perturbation space in  three ways.  In  Example 1 we look for the worst possible
perturbation, both locally and globally.  In Example 2 we add constraints to the perturbation space, representing prior
knowledge, and again look for maximally bad local and global perturbations. Finally, in Example 3, we marginalise over the
perturbation space -- rather than optimising over it --  as a way of dealing with the uncertainty of the prior. 

\bibliographystyle{abbrvnat}
\bibliography{jabref.bib}

\appendix

\section*{Appendix}\label{Appendix}

\begin{proof}[Lemma \ref{perturbed_post}]
\begin{eqnarray}
 \pi_p(\mu|x,\lambda)=\frac{\pi(\mu,\lambda) f(x;\mu)}{g(x,\lambda)}
\end{eqnarray}
where
\begin{eqnarray}
g(x,\lambda)&=&\int  \pi(\mu,\lambda;\theta) f(x;\mu) \,d\mu\nonumber\\
&=& \int f(x;\mu)\pi_{0}(\mu;\theta)\,d\mu +  \sum\nolimits_{j=2}^{k}{\lambda_j \int \, q_j(\mu,\theta) f(x;\mu)\pi_{0}(\mu;\theta)\,d\mu}\nonumber\\
&=& g(x) \left\{ 1 + \sum\nolimits_{j=2}^{k}{\lambda_j \, E_p^0[q_j(\mu,\theta)]}\right\}\label{marginal_perturb}
\end{eqnarray}
Since $f(x;\mu)\pi_{0}(\mu;\theta)=g(x) \pi_p^{0}(\mu|x,\theta)$ and $g(x)= \int f(x;\mu)\pi_{0}(\mu;\theta)\, d\mu$ 
where, $g(x)$ is the marginal density of sample in the base model. Hence, 
\begin{eqnarray}
\pi_{p}(\mu,\lambda|x;\theta) &=& \frac{  f(x;\mu)\pi_{0}(\mu;\theta) \left\{1 + \sum\nolimits_{j=2}^{k}{\lambda_j \, q_j(\mu,\theta)}\right\}  }{g(x) \left\{ 1 + \sum\nolimits_{j=2}^{k}{\lambda_j \, E_p^0[q_j(\mu,\theta)]}\right\}}\nonumber\\
&=&\frac{\pi_{p}^{0}(\mu|x,\theta)}{\xi(\lambda,\theta)} \left\{1+\sum\nolimits_{j=2}^{k}{\lambda_j \, q_j(\mu,\theta)}\right\},\hspace{.5cm} \lambda \in \Lambda_{\theta} \nonumber
\end{eqnarray}

with \hspace{3.5cm}$\xi(\lambda,\theta)=1 + \sum\nolimits_{j=2}^{k}{\lambda_j \, E_p^0[q_j(\mu,\theta)]}$.\\

Also $\xi(\lambda,\theta)>0$, since $h^*(\mu;\lambda,\theta)> 0$, for all $\mu \in
\mathbb{R}$ and $\lambda \in \Lambda_{\theta}$, and $\xi(\lambda,\theta)=E_p^0(h^*(\mu;\lambda,\theta))$.
\end{proof}

\begin{proof}[Lemma \ref{perturbed_moments}]
Result follows by direct calculation and using the fact that, 
\begin{eqnarray}
A_j^l(x):=\int{\mu^l q_j(\mu)\,\pi_{post}^{0}(\mu|x)\,d\mu}=E_{p}^{0}[\mu^l q_j(\mu)]\label{A-j^l}
\end{eqnarray}
\end{proof}

\begin{proof}[Theorem \ref{varphi_form}]
Substitute $u^*(\cdot)$ by $u(\cdot)$ in \citep[Result 8]{Gustafson1996}.
\end{proof}

\begin{proof}[Theorem \ref{Psi_theorem}]
 By direct calculation and use of equation (\ref{A-j^l})
\end{proof}

\begin{proof}[Lemma \ref{posterior_predict}]
\begin{eqnarray}
g_{p}(y)&=& \int f(y;\mu)\pi_{p}(\mu,\lambda|x)\,d\mu\label{convolv}
\end{eqnarray}
is the convolution of $\mathcal{N}(\mu,\sigma^2)$ and $\mathcal{N}(\mu_{\pi},\sigma^2_{\pi})$. Since, 
$$ \frac{(y-\mu)^2}{\sigma^2}+\frac{(\mu-\mu_{\pi})^2}{\sigma^2_{\pi}}=   \frac{\left(\mu-   \frac{\sigma_{\pi}^2 y+\sigma^2 \mu_{\pi}} {\sigma^2 +\sigma_{\pi}^2}\right)^2}{\frac{\sigma^2 \sigma_{\pi}^2}{\sigma^2 +\sigma_{\pi}^2}} + \frac{(y-\mu_{\pi})^2}{\sigma^2+\sigma_{\pi}^2}                                 $$
hence, the posterior predictive distribution for base model is $\mathcal{N}(\mu_{\pi}, \sigma_{\pi}^2+\sigma^2)$ 
and (\ref{g_new_final}) is obtained by direct calculation, where,
$$\Gamma= \frac{1}{\sqrt{2\pi (\sigma_{\pi}^2+\sigma^2)}}\exp\left\{ - \frac{(y-\mu_{\pi})^2}{2(\sigma_{\pi}^2+\sigma^2)}\right\}$$
and $E^{\star}(\cdot)$ is expectation with respect to $\mu$ according to the following normal distribution 
$$\mathcal{N}\left( \frac{\sigma_{\pi}^2y+\sigma^2 \mu_{\pi}}{\sigma_{\pi}^2+\sigma^2},  \frac{\sigma_{\pi}^2\sigma^2}{\sigma_{\pi}^2+\sigma^2} \right)$$
\end{proof}

\begin{proof}[Theorem \ref{Diverg_thorem}]
Use of Lemma \ref{posterior_predict} and direct calculation finishes the proof.  
\end{proof}

\begin{proof}[Lemma \ref{Manifold}]
Let $\sigma_0=1$ in equation (\ref{boundary plane}) for convenience and fix $\lambda_4$. From solving $P_{\lambda}(z)=0$ and
$P^{\prime}_{\lambda}(z)=0$, simultaneously for $\lambda_2$ and $\lambda_3$, we get a smooth parametrization for the boundary
as follows
\begin{eqnarray}
\left\{ \begin{array}{ll}
\lambda_2(z) = \frac{ \lambda_4\, (z^6-3z^4+9z^2+9)-3z^2+3 }{ z^4+3 } \\
\lambda_3(z) = \frac{ 2z \, [1-(z^4-2z^2+3)\lambda_4] }{z^4+3} \end{array} \right. \label{solution_set}
\end{eqnarray}
Hence, by implicit function theorem (\citealp[p.225]{Rudin1976}) the boundary of $\Lambda_{\theta_0}$ is a smooth surface (Manifold) embedded in $R^3$ by
\begin{eqnarray}
 \mathcal{S}_1 &:&R\times U \rightarrow V \nonumber\\
 \mathcal{S}_1 (z,\lambda_4)&=&[\lambda_2(z,\lambda_4),\lambda_3(z,\lambda_4),\lambda_4]\label{surf_1}
\end{eqnarray}

\end{proof}

\begin{proof}[Lemma \ref{direc_max}]
$\nabla \varphi=(a_2,a_3,a_4)$, is a vector originated at $\lambda=0$, where $a_j=Cov_p^{0}\left(\mu,q_j(\mu)\right)$.
If it is feasible then clearly gives the maximum direction. However, if it is not feasible then $a_4\leq 0$ since 
the condition $a_4>0$ is necessary for feasibility. Thus, the direction of the orthogonal projection of $\nabla \varphi$
onto the boundary plane corresponding to $\lambda_4=0$ is the closest we get to a maximum and feasible direction. 
\end{proof}

\end{document}